# Vector Boson Fusion Higgs Production at the LHC – Mass Variables

Dan Green
Fermilab

January, 2005

**Introduction – VBF**

The cross section for Higgs production at the LHC is shown in Fig. 1 below. The dominant process is gluon-gluon fusion into a virtual top loop which then radiates a Higgs. Nevertheless, the vector boson fusion (VBF) process where quarks radiate virtual W bosons which inverse W pair decay to form a Higgs, qqH, is a large cross section at all plausible Higgs masses.

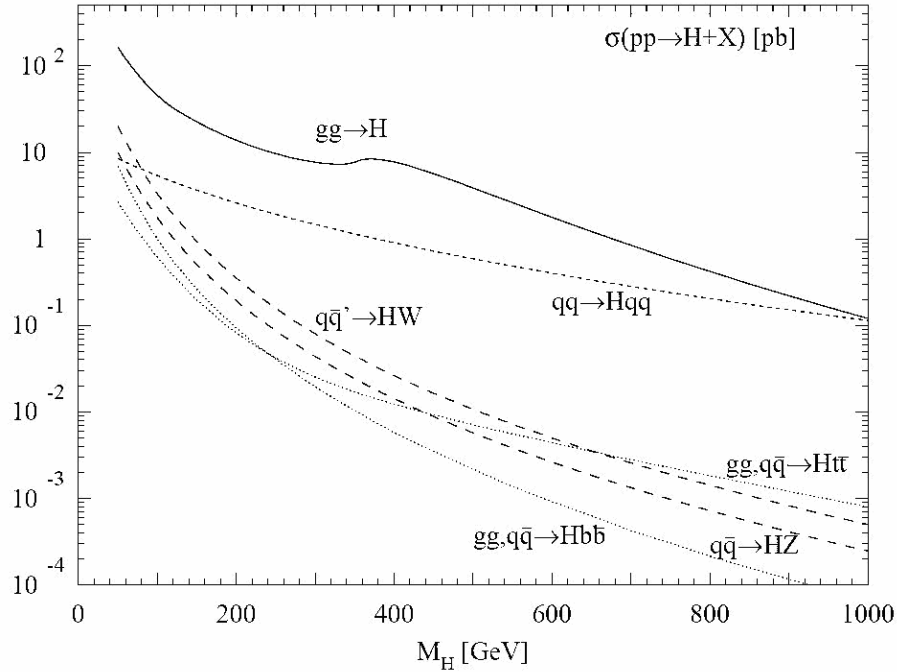

Figure 1: Cross sections at the LHC for Higgs boson production. Note that the VBF process is down a factor less than 10 with respect to the dominant gluon-gluon fusion process over the full possible mass range of the Higgs.

The VBF process is a favored discovery mode for a Higgs of mass greater than around 130 GeV because the two quarks which radiate the W pair which then fuse to make the Higgs continue in the forward,/backward direction and can be detected as "tag jets". The two small angle jets tag the emission of the two W's [1]. The existence of these two jets allows the experimenter to reduce the backgrounds that are present in more inclusive Higgs searches which exploit the dominant gluon-gluon fusion production process.

**Backgrounds**

The desired final state contains two "tag jets" at small angles to the proton beams and a Higgs at wide angles. The largest Higgs branching ratio over most of the mass range is into W pairs even at low mass when one W is somewhat off mass shell [1]. Indeed, this means that the process cross section depends only on the WWH coupling.

This fact by itself makes the WW plus two tag jet final state very worthwhile to study. Background levels and event statistics then imply that leptonic W decays are favored. This process will pass most trigger menus in that there are one or two leptons above some transverse momentum threshold, substantial missing energy taken off by the neutrinos and two moderate transverse momentum jets, $|\vec{P}_{tag}|_T \sim M_W/2$, near the beam directions. This event signature is sufficiently distinct that it can be efficiently triggered on.

One obvious background is the strong production of top quark pairs. The b pairs from the decays need only mimic the tag jets to make this process a major background. The cross section for this process is very large with respect to the Higgs, as seen in Fig. 2.

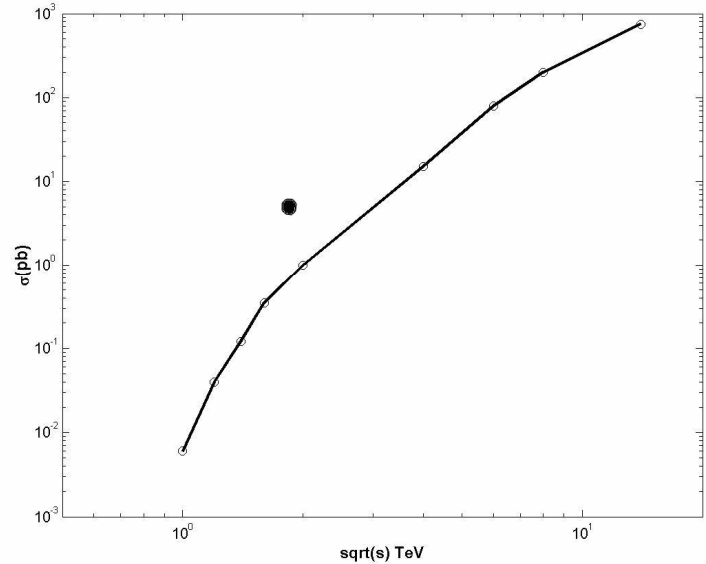

Figure 2: Cross section for the production of top quark pairs as a function of C.M. energy for p-p collisions. The dot indicates the Tevatron data point.

The Tevatron data point, for proton-antiproton production, gives some confidence in the extrapolation to the LHC. At the LHC the cross section is ~ 1000 pb, about 300 times larger than the qqH cross section for a Higgs boson of 200 GeV mass. Fortunately, it has been shown [1] that this background can be reduced to a negligible level. A more difficult background, indeed the dominant one, is the production of top pairs accompanied by a radiated gluon. The Feynman diagrams at tree level are shown in Fig.3. In the case of top quark pairs by themselves there are only three diagrams for s channel g + g -> g* with g* decay into top pairs or t channel top exchange.

The cross section for this process, labeled as ttg, evaluated in COMPHEP, exceeds the cross section for top pairs alone. At tree level the latter process proceeds through g + g into s channel g splitting into top pairs or by the t channel exchange of a gluon with subsequent splitting into a top pair. Clearly there are other processes available in the case of the ttg final state which explains why the cross section exceeds that for exclusive top pairs, labeled tt , even with the penalty paid of an additional strong coupling constant in the cross section for ttg ( see Table 1).

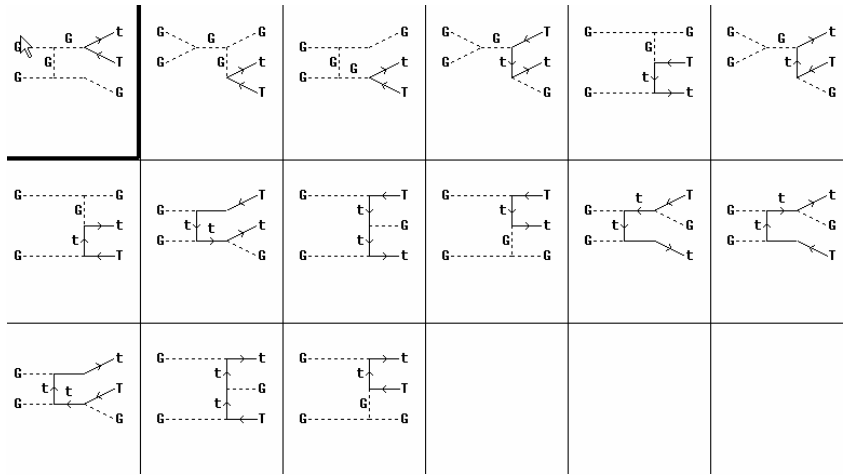

Figure 3: Feynman diagrams from COMPHEP on the production of a top quark pair accompanied by a gluon.

Another background is the non-resonant production of W pairs accompanied by two gluon jets which mimic the qqH process tag jets. Recently the predicted cross section for WW [2] has been confirmed by the Tevatron data taking [3] as indicated in Fig. 4.

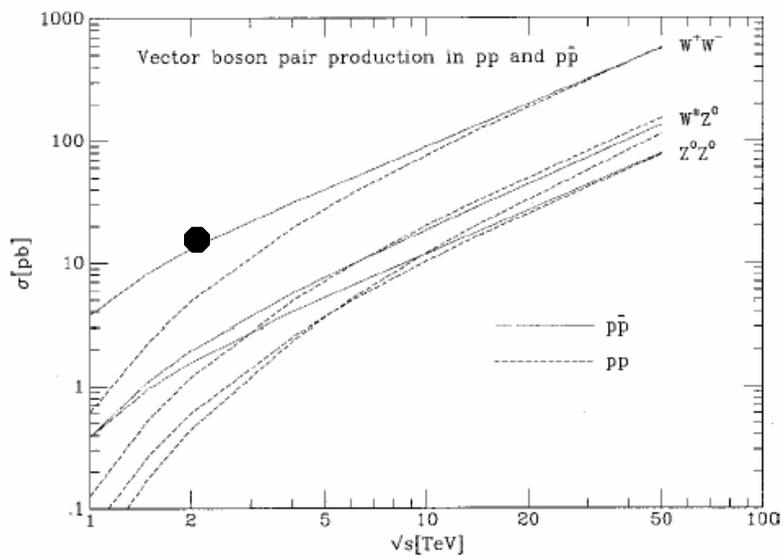

Figure 4: Cross section as a function of C.M. energy for the production of W pairs in proton-proton and proton-antiproton collisions. The dot indicates the Tevatron data point.

The agreement of Tevatron data with prediction gives some confidence in the extrapolation to the LHC. The related process with an additional two radiated gluons has a cross section not much reduced from that for the WW process. The tree level cross sections as evaluated using COMPHEP are shown in Table.1.

Table 1: Cross Sections for VBF H Production and Backgrounds

| Process | Cross Section (pb) | "Tag" Cuts |
|---|---|---|
| qqH | 3 | ~ 3 |
| tt | 700 | |
| ttg | 1800 | ~ 12 |
| WW | 80 | |
| WWgg | 25 | |

The major remaining background after cuts is the ttg process. It has been shown that the tt and WWgg processes can be reduced well below the remaining cross section for ttg [1, 4]. That assertion has been checked using the COMPHEP Monte Carlo program to generate all the processes shown in Table 1 and to then impose cuts on the jets which are forced to mimic the qqH tag jets ("tag" cuts). In what follows, the ttg process is considered the dominant background and the others are ignored.

**COMPHEP Evaluation**

The COMPHEP program was used to generate the processes shown in Table 1. The subsequent decays were assumed to be isotropic because COMPHEP performs a spin sum so that polarization information is lost. In future studies other programs will be used in order to explore the correlations between decay products induced by spin effects.

The qqH process was modeled at tree level with full matrix element evaluation rather than using the effective W approximation. Since the emitted W bosons are soft, there is a strong correlation between the longitudinal momenta of the initial state quarks and the tag jets, as shown in Fig. 5.

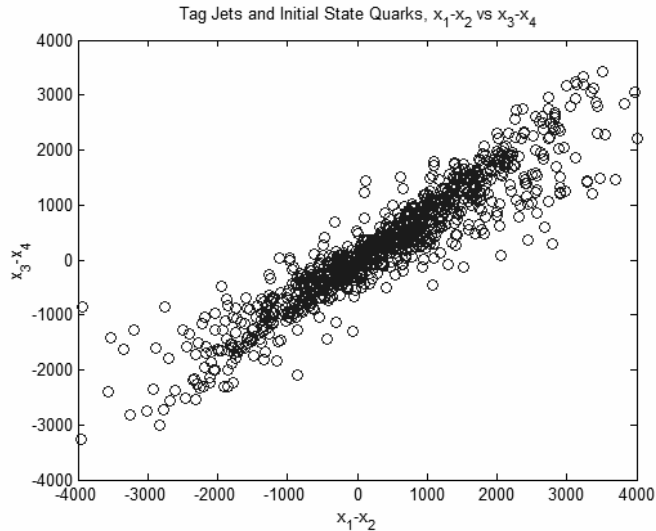

Figure 5: Correlation in the qqH process between the initial state quark longitudinal momentum difference and the final start tag jet momentum difference.

In contrast to the situation for inclusive Higgs searches, the transverse momentum of the Higgs can be determined by the measurement of the transverse momenta of the two tag jets. A check of the longitudinal momentum of the Higgs shows a rather weak correlation to the tag jet longitudinal momentum because the initial state does have a non-zero value of $x_1 - x_2$ (see Fig. 5). This is in contrast to the transverse momentum, where the initial state is known to have none on the scales considered here.

Therefore, if we wish to estimate the Higgs longitudinal momentum we must look elsewhere. The correlation of the final lepton pair momentum and the Higgs momentum was examined and is shown in Fig. 6 ( $H \rightarrow W^- + W^+ \rightarrow \ell^- + \bar{\nu}_\ell + \ell^+ + \nu_\ell$ ).

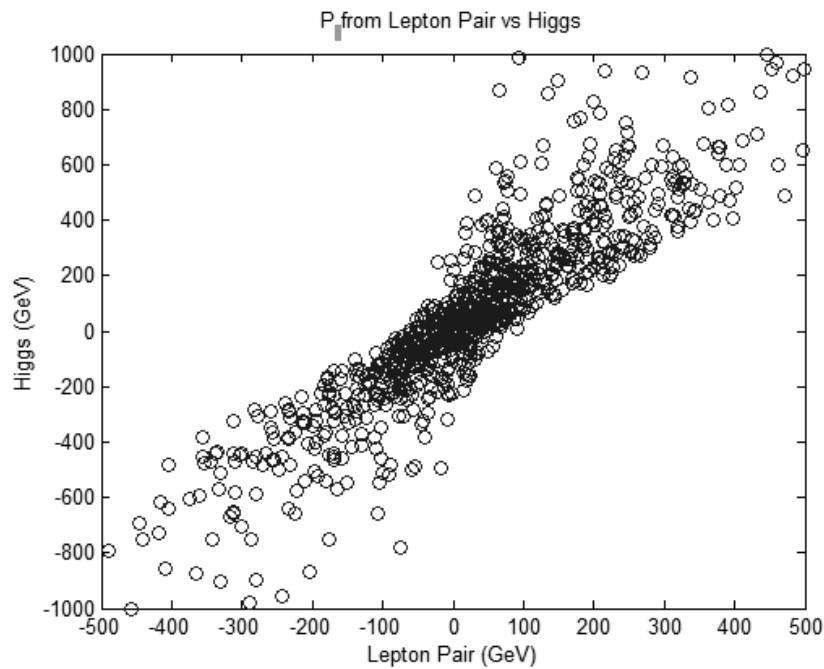

Figure 6: Correlation in the qqH process of the longitudinal momentum of the lepton pair and the Higgs. Decays H -> W + W -> l + v + l + v are modeled assuming isotropy.

Clearly, there is a fair degree of correlation between the lepton pair and the parent Higgs. Therefore, the Higgs longitudinal momentum is approximated as being twice the lepton pair longitudinal momentum.

For the ttg background the top quarks were modeled to decay isotropically into W + b. The g was required to mimic one tag jet with transverse momentum above 5 GeV and with rapidity greater than 1 and less than 5. One of the b was required to mimic the other tag jet and have a rapidity difference of greater than 2.5 compared to the gluon. Therefore, the g-b are the tag jet backgrounds. There is an additional b and it is required to have small transverse momentum, less than 20 GeV. This generous "veto" is a crude attempt to allow for pileup minimum bias events faking a soft jet in the central region of the event, |y| < 3. In reality a lower value of the cut may be possible, but one should wait for data before 'fine tuning" the analysis.

After these simple and soft cuts the ttg events remain four times larger in cross section than the qqH events, as seen in Table 1. Clearly, additional cuts are needed. In previous studies [1] the Higgs mass parameter was chosen to be the transverse mass. This quantity is distributed essentially uniformly from zero to the Higgs mass. The background is also distributed roughly uniformly although it extends to higher values of the transverse mass. Therefore, rather than searching for a resonant Higgs peak, one is reduced to doing a "counting experiment". However, a counting experiment requires a very good understanding of the detector which is used and a good understanding of the systematic uncertainties of the cross sections of all the backgrounds with regard to trigger, reconstruction and cut efficiencies.

In fact, to reduce the uncertainties, the signal to background ratio can be increased by imposing additional cuts. However, the most effective cuts, [1, 4], are ones which assume that the search is restricted to a scalar particle, as illustrated in Fig. 7. Assuming a spin zero parent, the W spins are paired oppositely and then the V-A nature of the weak W decays then implies that the leptons have roughly the same direction. This kinematic correlation is the origin of the cuts [1], [4] imposed on the leptons – small invariant pair mass and small relative azimuthal angle.

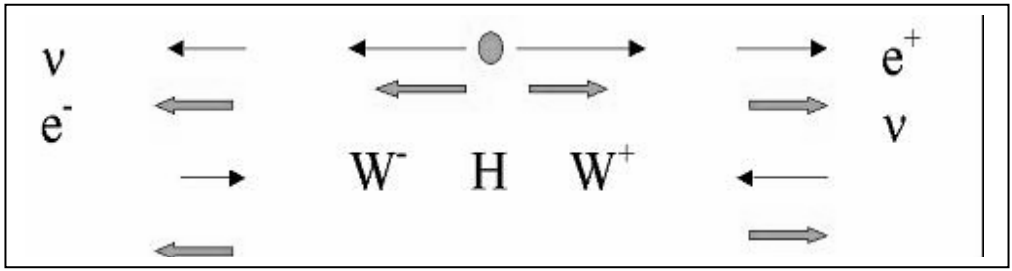

Figure 7: Kinematic relations induced by the spin zero nature of the Higgs and the V-A helicity structure of the W decay into lepton pairs. Spin direction is indicated by thick arrows, spatial direction by thin arrows.

**Additional "Tag" Jet Cuts**

Rather than search solely for a scalar resonance in a counting experiment, a search without assuming a parent spin and using a resonant peak was looked for. First, the rapidity difference can be cut harder, at the expense of statistical significance. The rapidity differences for signal and background are shown in Fig. 8. Clearly one can see if a peak exists in the rapidity difference distribution. If it can be observed, a relatively clean VBF event sample has been obtained.

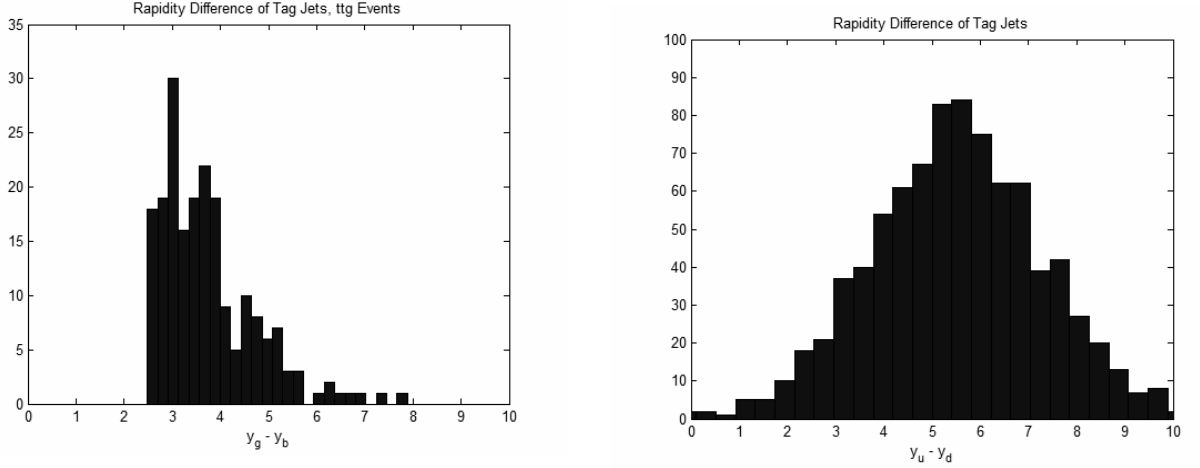

Figure 8: "Tag jet" rapidity differences for ttg and qqH events respectively.

The real tag jets have another difference from the mimics. They tend to have larger transverse momentum which can be used to cut harder on the mimic tag jets. There are also differences in the mass of the tag jet pair [1, 4]. A tag jet longitudinal mass parameter is defined in Eq.1, assuming massless jets.

$$M_{tag}^2 = (|\vec{P}_{tag1}| + |\vec{P}_{tag2}|)^2 - (\vec{P}_{tag1} + \vec{P}_{tag2})_{\parallel}^2 \quad (1)$$
$$M_{tag} - 3.5 |\vec{P}_{tag1} + \vec{P}_{tag2}|_T$$

The distribution of this mass parameter for qqH and ttg events is shown in Fig.9.

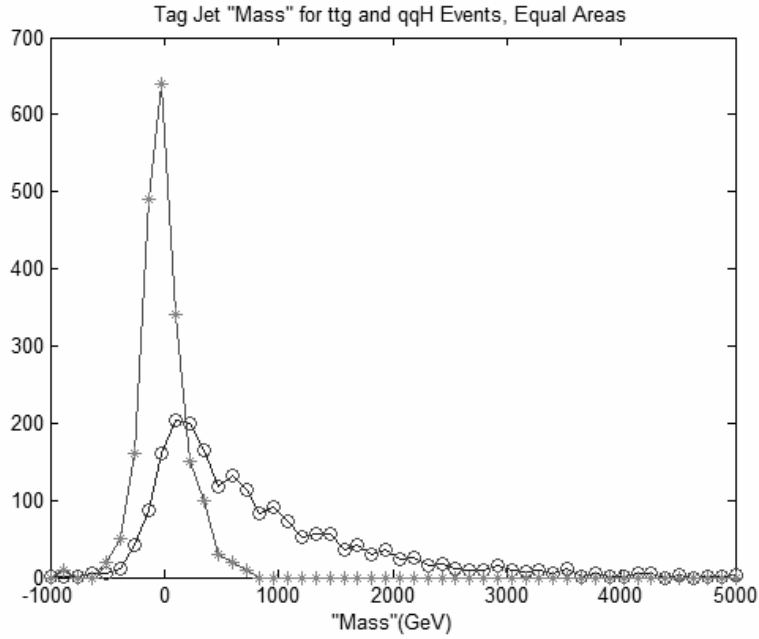

Figure 9: Tag jets "mass parameter" defined in Eq. 1, for qqH ( o ) and ttg ( * ) events normalized to equal cross sections.

Clearly, a rather clean separation of signal and background is possible if a roughly 500 GeV cut is made on this parameter. The existence of additional cuts appears to be promising, although they come at the expense of reduced statistics. Clearly, a full simulation including detector and fragmentation effects is needed before the efficacy of these cuts is really established.

**Rapidity Ordering and Tag-Lepton Correlations**

The correlations between the tag jets have been examined and yield useful mass cut parameters. There are also correlations between the tag jets and the leptons arising from the decay of the two produced W particles. We first look at angular ordering. The tag jet with the largest positive rapidity is called the leading +y tag jet, while the tag with the largest negative rapidity is called the trailing –y tag jet. The leading lepton has the largest positive rapidity value. As seen in Fig. 10, the lepton trails the leading +y tag jet for the qqH events, while this angular ordering is not observed for about half the ttg background events.

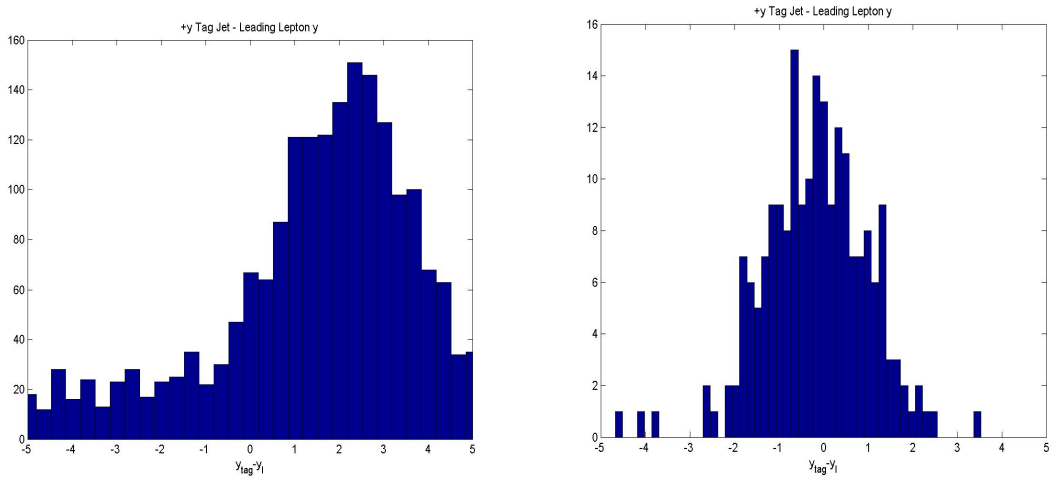

Figure 10: Rapidity difference between the tag jet and the lepton in the +y leading case for qqH events (left) and ttg events (right).

There are also differences due to transverse variables. The mass of the leading tag jet and the leading lepton was computed, along with the mass of the trailing tag jet and the trailing lepton. In both cases the mean mass is larger for the qqH events rather than the ttg events. Plotted in Fig. 11 is the square root of the product of the two masses for qqH and ttg events, normalized to equal number of events after the ttg events are cut at a rapidity difference of 2.5 and one of the b quarks is "vetoed" by requiring a transverse momentum less than 20 GeV. The mean of this mass parameter is 107 GeV for ttg events and 193 GeV for qqH events. Clearly, this cut is also available to increase the signal to background ratio.

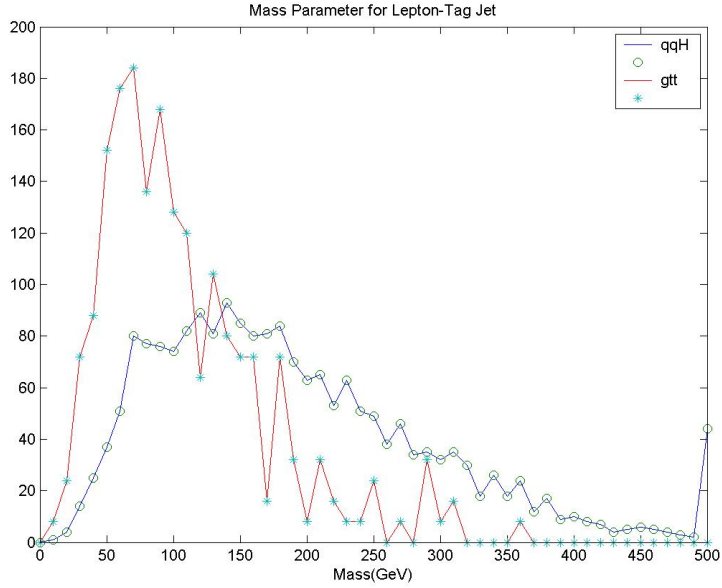

Figure 11: Tag jet- lepton mass parameter for qqH and ttg events normalized to equal numbers of events.

**Higgs "Mass" Parameter**

Finally, as discussed above, the Higgs momentum-energy components can be estimated as shown in Eq.2. The particular choice of components was chosen such that the squared mass parameter is always positive. The transverse momentum magnitude is subtracted simply to make the mass scale be roughly the true mass scale. That this parameter changes with Higgs parent mass was checked at 180 and 200 GeV. Other choices are clearly possible.

$$\begin{aligned}
E_H &= 2(E_{\ell 1} + E_{\ell 2}) + \slashed{E}_T \\
(\vec{P}_H)_T &= -(\vec{P}_{tag1} + \vec{P}_{tag2})_T \\
(\vec{P}_H)_\parallel &= 2(\vec{P}_{\ell 1} + \vec{P}_{\ell 2})_\parallel \\
M_H^2 &= E_H^2 - |\vec{P}_H|_T^2 - |\vec{P}_H|_\parallel^2 \\
M_H &- (\vec{P}_H)_T
\end{aligned} \qquad (2)$$

The distribution of this mass parameter for both qqH and ttg events is shown in Fig. 12. Clearly, the goal of making a "mass bump" search for a resonant structure rather than a "counting experiment" has been roughly achieved. The mass parameter has, however, a rather large width, which limits the utility of this parameter. Indeed, the behavior of the shape for signal and background when additional cuts are imposed needs to be studied.

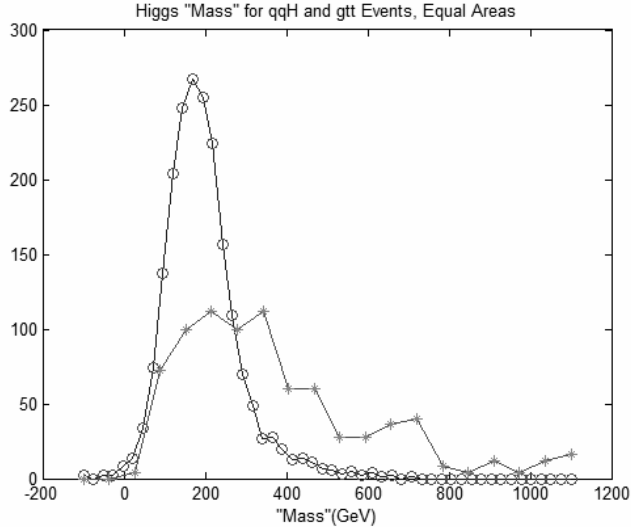

Figure 10: Higgs "mass parameter" defined in Eq. 2, for qqH ( o ) and ttg ( * ) events normalized to equal cross sections.

However, the background rises smoothly from near zero mass but rather more slowly than the signal. One can hope that the different behavior for signal and background is maintained when more detailed studies are performed in the near future.

**qqZ**

There are several predictions for W and Z plus a number of additional jets [5]. Again, Tevatron data points give confidence in the extrapolation to the LHC. Indeed, the Feynman VBF diagrams for this process are identical to those for VBF production of the Higgs. However, the cross section for qqZ is much larger than that for qqH. In addition, these events will also appear in the same di-lepton plus forward/backward jets trigger stream as the qqH events. Therefore, long before the qqH events become statistically significant, the VBF mode can be studied without first imposing cuts on the tag jets because the leptons supply the trigger [6]. In addition, a sharp Z resonance in the dilepton mass spectrum is available to pull the qqZ VBF events out of any residual background.

Therefore, the qqZ process will supply an invaluable initial look at VBF and contribute greatly to the evolving strategy to use VBF in searching for a new resonance in the WW system. In particular, the tag jet properties may be studied prior to the distortions imposed by making harsh cuts. If so, then cuts well tailored to the experimentally observed VBF tag jets can be fashioned.

**References**

1) N. Kauer, T. Plehn, D. Rainwater and D. Zeppenfeld, Fermilab Pub 00/322 T and arXiv:hep-ph/00121351, Dec. 27, 2000
2) J.M. Campbell and R.K. Ellis, Phys. Rev. D 60, 113006 (1999)